\RequirePackage{fix-cm}
\documentclass[smallextended]{svjour3}       % onecolumn (second format)
\smartqed  % flush right qed marks, e.g. at end of proof
\usepackage{graphicx}
\usepackage{amsmath}
\usepackage{amssymb,amsmath}
\usepackage{subfigure}
\usepackage{anysize}

\numberwithin{equation}{section}

\renewcommand{\epsilon}{\varepsilon}

\journalname{Theoretical Ecology}
\begin{document}
\renewcommand{\thefootnote}{\fnsymbol{footnote}}

\title{Synchronization in ecological systems by weak dispersal coupling with time delay}\thanks{%Grants or other notes
}


\author{Emily Wall \and Frederic Guichard \and Antony R.\ Humphries}


\institute{E.\ Wall \and F.\ Guichard \at
               Department of Biology, McGill University, Montreal, Canada\\\
%Stewart Biology Building
%1205 Docteur Penfield
%Montreal, Quebec
%Canada H3A 1B1
%             Tel.: +123-45-678910\\
%             Fax: +123-45-678910\\
              \email{emily.wall@mail.mcgill.ca}            \\
%           F.\ Guichard \at
%               Department of Biology, McGill University, Montreal, Canada\\
           \and
           A.R.\ Humphries \at
               Department of Mathematics and Statistics, McGill University, Montreal, Canada\\	
%Burnside Hall, Room 1005,
%805 Sherbrooke Street West,
%Montreal, Quebec,
%Canada H3A 0B9
}

\date{Received: date / Accepted: date}

\maketitle

\begin{abstract}
One of the most salient spatio-temporal patterns in population
ecology is the synchronization of fluctuating local populations across
vast spatial extent. Synchronization of abundance has been widely
observed across a range of spatial scales in relation to rate of
dispersal among discrete populations.
However, the dependence of
synchrony on patterns of among-patch movement across
heterogeneous landscapes has been largely ignored.
Here we consider the duration of movement
between two predator-prey communities connected by weak dispersal, and
its effect on population synchrony. More specifically, we introduce
time delayed dispersal to incorporate the finite transmission time
between discrete populations across a continuous landscape. Reducing
the system to a phase model using weakly connected network theory, it
is found that the time delay is an important factor determining the
nature and stability of phase-locked states. Our
analysis predicts enhanced convergence to stable synchronous
fluctuations in general, and a decreased ability of systems to produce in-phase synchronization dynamics in the presence of delayed dispersal. These results
introduce delayed dispersal as a tool for understanding the importance of
dispersal time across a landscape matrix in affecting metacommunity dynamics. They further
highlight the importance of landscape and dispersal patterns for
predicting the onset of synchrony between weakly-coupled populations.

\keywords{Synchronization \and Phase model \and Dispersal \and Time delay }
\end{abstract}

\section{Introduction}\label{intro}
Predicting the onset and maintenance of synchronous fluctuations of
abundance among populations has led to important progress in the
understanding of population persistence, species coexistence, and
response to environmental change. However, most theoretical studies of
synchrony have assumed instantaneous and passive movement of
individuals among discrete habitats. In natural systems, discrete
populations are typically embedded within a landscape matrix that
constrains patterns and duration of movement between populations. It is
thus important to predict the impact of such a landscape matrix on
the emergence and stability of synchronous fluctuations of abundance
between distant populations. Here we incorporate delayed dispersal as a
means to study landscape and dispersal patterns between discrete
predator-prey communities.

Synchronization is generally understood as a phenomenon where through interactions with one another, the
timings of fluctuations in components of a dynamical system lock to
a single pattern of variation in time. This concept arises in a variety of systems in
physics, chemistry, social sciences, and biology: from crickets
chirping in synchrony, to fireflies able to all flash in unison
(Blasius and T\"{o}njes 2007). Spatial synchronization in the context of
population ecology refers to coincident changes in population
characteristics such as abundance, reproduction, mortality, or mean
size or age distribution, in geographically separated populations
(Liebhold et al. 2004). Ecological examples of synchronization include
the cycling of hare-lynx populations, where fluctuations of groups across Canada were found to
be locked on the same tight ten-year cycle
(Blasius and T\"{o}njes 2007). Synchronized population dynamics have been observed in
natural systems across a variety of taxa and spatial scales, from fungal plant pathogens across 0.5-3m, insect
herbivores across 1-1000km, to birds across 5-2000km
(Liebhold et al. 2004).

 For our purpose we study the synchrony of coupled oscillators, where individual populations undergo intrinsic oscillations in abundance with some natural frequency, and adjust their oscillations due to coupling (dispersal) between oscillators. This formalism has been used to predict the onset of synchrony in simple ecological systems (Goldwyn and Hastings 2008, Goldwyn and Hastings 2011, Bresloff and Lai 2012). Networks of
time-delayed and weakly connected oscillators have also been well-studied
in physics, engineering, and neurophysiology (Schuster and Wagner 1989, Dhamala et al. 2004, Campbell and Kobelevskiy 2012). However, no
study has elucidated the role of delayed dispersal in coupled
ecological systems.

Causes of spatial synchrony have been studied extensively across
systems and scales. Coupled oscillator theory predicts that
oscillators can be synchronized through direct coupling, or by external forcing. Equivalently in ecology, there are two main
mechanisms that produce synchrony: $(a)$ density-dependent direct
interactions, namely dispersal between populations, and $(b)$ what is
known as the Moran effect  (Moran 1953), which is the entrainment
of systems with similar density-dependent dynamical structures by
correlated density-independent external factors, such as
climate or weather (Liebhold et al. 2004). Because most natural systems
are connected through both dispersal and external stochastic
fluctuations, and because both mechanisms produce similar patterns of
synchrony it is often difficult to determine which factor or
combination of factors underlies observed patterns of spatial
synchrony.

Theory predicts that populations oscillating due to the same or
slightly different density-dependent processes (linear or nonlinear)
can be synchronized by dispersal of just a very small number of
individuals per generation (Liebhold et al. 2004). However, if
density-dependent processes are so different that populations
oscillate with very different frequencies, then synchronization may
not be possible through dispersal (Ranta et al. 1998). Besides dynamical stability of synchronization fluctuations, the rate of synchronization is another
constraint on the ability of dispersal to explain synchrony in natural
populations: dispersal-induced synchrony must converge rapidly to be observed (Goldwyn and Hastings 2008). Otherwise, populations will be kept in
transient non-synchronous states by environmental perturbations. While
strong coupling (i.e., a large number of individuals disperse per
generation) can always increase local convergence rates to synchrony in
general, in a two predator-prey patch system connected only weakly by
dispersal, the required property for fast convergence to synchrony is
the separation on time scales between predator and prey dynamics (Goldwyn and Hastings 2008). This property is characteristic of relaxation oscillators where a large part of the cycle is spent at low
population densities. In contrast, sinusoidal oscillators take a very long time to synchronize
by weak dispersal. Our study examines some
mechanisms underlying this result and uses its robustness to
delayed dispersal to further predict characteristics of ecological systems that favour weak dispersal-induced synchronization.

Metapopulation theory and the study of fragmentation have emphasized
discrete boundaries and movement between populations (Levins 1969, Hanski 1999). The
appeal of metapopulation theory lies in its simplicity associated with
the assumption of passive and instantaneous population
movement between patches. As a result, model predictions applied to natural systems usually ignore the complex landscape matrix that provides the context
for discrete populations (Brady et al. 2011, Turner et al. 2001): mountain peaks embedded in
valleys, forest patches within meadows and seagrass between coral
reefs. Dispersing propagules and migrating individuals can spend
significant time and show non random movement within such a landscape
matrix depending on the nature of the movement and on distance between populations. Current
metapopulation theory assumes that the spatial structure of
metapopulations can be captured by a per capita dispersal rate alone, rather than dispersal time.

The study of spatial synchrony is important because of the implications of synchronous fluctuations of abundance on global population persistence: populations oscillating in full synchrony are more susceptible to extinction at low population density because of reduced `rescue effect' (Brown and Kodric-Brown 1977) from neighbouring patches. Population asynchrony has been associated with increased global population persistence in experiments (Huffaker 1958; Holyoak 2000).	

To better understand the mechanisms leading to synchronization in
natural populations, we study the role of weak and delayed dispersal
in driving the ecological conditions for the existence and stability of synchronous
states, and for their local convergence rate. We
illustrate how time delayed dispersal can be used to implement the
particulars of individual movement (through active or passive
dispersal) in the space between patches. While discrete patch models
are useful for studying synchronization of spatially
distributed systems from a dynamical systems point of view, they are
associated with the assumption of instantaneous dispersal where
individuals disperse with no transmission delay. By including a time
delay that retains more detailed information about the dispersal
process, we extend the patch dynamics model to integrate properties of the
landscape matrix that contains discrete populations.

We first
formulate an ecological model as a two-patch Rosenzweig-MacArthur predator prey model
coupled through prey dispersal with a time delay. We reduce our model to a phase model using weakly-coupled network theory (Appendix~A).
Our results show that
synchronization through weak dispersal is
highly sensitive to dispersal time. We find that with slight
variation in the delay, the overall influence of weak dispersal on the
rate of convergence to synchrony can vary greatly. This means that
dispersal time between spatially discrete systems is a key
variable for assessing the role of weak dispersal as a cause of
synchronous fluctuations in natural systems. We further extend
recent results from Goldwyn and Hastings (2008) on synchronization between patches
to predict the importance of such a matrix on the synchronization of
weakly coupled predator-prey metacommunities.

\section{Ecological model}\label{eco}
Following Goldwyn and Hastings (2008), we use the
Rosenzweig-MacArthur predator-prey model for single-patch dynamics:
\begin{align}
\begin{split}
\frac{dH}{dt}&=rH\Bigl(1-\frac{H}{K}\Bigr)-\frac{acPH}{b+H},\\
\frac{dP}{dt}&=\frac{aPH}{b+H}-mP. \label{single}
\end{split}
\end{align}
Here $H(t)$ and $P(t)$ represent the size of the prey and predator
populations respectively. Prey growth is modeled by logistic growth
with intrinsic rate $r$ and carrying capacity $K$, and the intake of
prey by predators is modelled with a Holling Type II functional
response specified by the parameters $a$ and $b$ and where the
conversion ratio of loss of prey to increase in predators is
$c>1$. The predators have a linear mortality rate, with parameter $m$.

Now suppose we have two identical and spatially discrete predator-prey patches, each with dynamics following (\ref{single}). We consider the case where only prey disperse between patches with a fixed dispersing rate $D$. The dispersing prey take some finite transmission time $s$ to cross the space between the patches, and this parameter can be derived using from details of individual movement during dispersal (see Appendix~D for details). Our delayed-dispersal predictions are compatible with previous theories of time delayed dispersal in two patch models (Prasad et al. 2008), and are shown to more broadly integrate existing spatial ecological frameworks for spatial dynamics (see Appendix~D).

If we further assume the absence of growth or mortality during dispersal, we can equate the incoming flux of individuals to a patch at time $t$ with the outgoing flux of individuals at the opposite patch at a time $t-s$, before the present time. This assumption allows us to incorporate the dispersal process into our four-equation model (predator and prey populations in 2 patches) with a simplified prey dispersal term $D(H_j(t-s)-H_i(t))$ in patch $i$. This assumption of no birth or mortality during dispersal is not necessary for our analysis in this section or in Section~\ref{theo}; however, it is crucial for our numerical computation of the function $H$  (Section~\ref{analysis}).

The result is the following two-patch Rosenzweig-MacArthur predator-prey model coupled through prey dispersal with a time delay:
\begin{align}
\begin{split}
\frac{dH_i}{dt}&=rH_i\Bigl(1-\frac{H_i}{K}\Bigr)-\frac{acP_iH_i}{b+H_i}+D\bigl(H_j(t-s)-H_i(t)\bigr),\\
\frac{dP_i}{dt}&=\frac{aP_iH_i}{b+H_i}-mP_i, \qquad i,j=1,2;\:\; i\neq{}j. \label{original}
\end{split}
\end{align}
This is identical to the model studied by Goldwyn and Hastings (2008), except
for the time delay of $s$ time units that we include in the dispersal term, and
that we allow only one of the species to disperse between patches.

We nondimensionalize (\ref{original}) following Goldwyn and Hastings (2008) to
reduce the number of parameters in a way that will be useful later for
adjusting the timescale separation of the predator and prey growth
rates. Thus we work with the following resulting equivalent form:
\begin{align}
\begin{split}
\frac{dh_i}{dt}&=\frac{1}{\epsilon}\left(h_i(1-\alpha h_i)-\frac{p_i h_i}{1+h_i}\right)
+d\bigl(h_j(t-\tau)-h_i(t)\bigr),\\
\frac{dp_i}{dt}&=\frac{p_i h_i}{1+h_i}-\mu p_i, \qquad i,j=1,2;\:\; i\neq{}j,
\end{split} \label{nondim}
\end{align}
where we reuse the variable $t$ for scaled time $at$, and make the
following substitutions:\\
$h_i=H_i/b$, $p_i=[ac/rb]P_i$, $\tau=as$, $\alpha=b/K$, $\mu=m/a$, $\epsilon=a/r$, $d=D/a$.

The single-patch dynamics of the Rosenzweig- MacArthur model are
well-understood (Goldwyn and Hastings 2008). There is a region of
$\epsilon$--$\alpha$--$\mu$ parameter space in which the dynamics produce a
stable limit cycle, namely: $\alpha<1$ and
$\mu<\frac{1-\alpha}{1+\alpha}$. This is the case we are interested in
for studying synchronization dynamics.

Furthermore, we are concerned with the part of
$\epsilon$--$\alpha$--$\mu$
parameter space where the predator-prey oscillations are
relaxation-like, since this leads to faster convergence to synchrony
when the patches are weakly coupled, a requirement for weak dispersal
to be the cause of synchronization in nature. Goldwyn and Hastings (2008)
explains how the separation of timescales between the predator and
prey leads the single-patch system to relaxation-like oscillations: in
the nondimensionalized form (\ref{nondim}), this is achieved by
decreasing any one of $\epsilon$, $\alpha$, or $\mu$. Decreasing $\epsilon$ increases the rate of intrinsic prey growth relative to that of the predator; decreasing
$\alpha$ increases the carrying capacity for prey, enhancing the size
and time between prey outbreaks; decreasing $\mu$ means slower
predator mortality, which leads to longer times for the predator
population to decrease to the point where the prey population spikes,
so that these spikes happen less frequently (Goldwyn and Hastings 2008). In
effect, the prey populations spend more time at low numbers and then
grow more rapidly at spikes when $\epsilon$, $\alpha$, and $\mu$ are
small.

\section{Phase model}\label{theo}
While non-delayed phase model reduction has been used in previous studies of predator-prey dynamics (Goldwyn and Hastings 2008, Goldwyn and Hastings 2011), our implementation of weakly connected network theory to reduce a time-delayed predator-prey system to a phase model is new. The only assumptions required are that coupling between stable limit-cycle oscillators is weak (i.e., $\delta$ is small; we follow Goldwyn and Hastings 2008 and assume $\delta=0.001$ corresponds to weak coupling), and that the explicit time delay $\tau$ is on the order of magnitude of the period of the oscillation, or less. In other words, the delay must not be larger than roughly 10 times a period of oscillation. The latter assumption is likely to apply to natural systems, where dispersal time is limited by individual lifespan.

Suppose we choose parameters $\epsilon$, $\alpha$, and $\mu$ for our
model (\ref{nondim}) so that the identical uncoupled subsystems (i.e., the two patches with $\delta=0$) oscillate on an exponentially orbitally stable limit cycle
$\gamma(t)$, which means that solutions starting
close enough to the limit cycle approach it exponentially fast in the
limit as time goes to infinity. The coupling in (\ref{nondim}) is symmetric, so  the system can be written in the general form for weakly connected networks as follows:
\begin{align*}
\dot{X}_1(t) &=F(X_1(t))+\delta W(X_1(t),X_2(t-\tau))\\
\dot{X}_2(t) &=F(X_2(t))+\delta W(X_2(t),X_1(t-\tau)),
\end{align*}
where
\begin{align}
\begin{split}
&X_i=(h_i,p_i)^T\!,\\
&F(X_i)=\left(\frac{1}{\epsilon}\Bigl(h_i(1-\alpha h_i)-\frac{p_i h_i}{1+h_i}\Bigr), \frac{p_i h_i}{1+h_i}-\mu p_i\right)^T\!,\\
&W(X_i(t),X_j(t-\tau))=(h_j(t-\tau)-h_i(t), 0)^T\!. \label{full}
\end{split}
\end{align}

Characterized by its shape, position and natural frequency $\Omega$,
each periodic solution $\gamma$ is an isolated closed curve through two-dimensional phase space, a distorted circle that can be parameterized with a phase variable $\theta$ that increases by $2\pi$ in one period
(see Appendix~A).

When we consider the full system, the fact that the $\gamma$ are
exponentially orbitally stable limit cycle attractors means that for
small $\delta$, coupling between the $X_i$ only significantly affects
the phase variables $\theta_i$. Our other required assumption is that the time delay $\tau$ is of an order of magnitude of $2\pi/\Omega$ or less. Then, by the theorem in Appendix~A we can reduce (\ref{full}) to the corresponding two-dimensional phase model:
\begin{align}
\begin{split}
\frac{d\theta_1}{dt}&=\Omega+\delta{}H(\theta_2-\theta_1-\Phi)\\
\frac{d\theta_2}{dt}&=\Omega+\delta{}H(\theta_1-\theta_2-\Phi),
\end{split}\label{2d}
\end{align}
where $\Phi=\Omega\tau \mbox{ mod }2\pi$ and
\begin{equation}
H(x)=\frac{1}{T}\int_0^T\hat{\gamma}(t)\cdot W(\gamma(t), \gamma(t+x/\Omega))\,dt, \label{H}
\end{equation}
with $\hat{\gamma}(t)$ solving the system
\begin{align}
\begin{split}
\frac{d\hat{\gamma}(t)}{dt}&=-DF(\gamma(t))^T \hat{\gamma}(t) \\
&\hat{\gamma}(t)\cdot \gamma{'}(t)=1.\label{adjointz}
\end{split}
\end{align}

Importantly, we see that the explicit time delays present in (\ref{full}) appear in the phase model as simple phase shifts in the coupling functions, which are more mathematically tractable than differential delay equations. Briefly (see Hoppensteadt and Izhikevich 1997 for full proof), this works because time-delayed terms in the phase model approximation to (\ref{full}) show up on the order of $\delta^2$, which we consider negligible in weakly connected systems.

Finally, we define the phase difference variable $\phi(t)$ as
$\phi\equiv\theta_1-\theta_2$. This reduces the system (\ref{2d})
further, to
\begin{equation}
\frac{d\phi}{dt}=G(\phi):=\delta[H(-\phi-\Omega\tau)-H(\phi-\Omega\tau)]\equiv \delta H_{delay}(\phi) \label{1d}
\end{equation}
where we just denote $\Phi$ with $\Omega\tau$.

\section{Phase model analysis}\label{analysis}

So far we have reduced the full system (\ref{full}) to a
one-dimensional dynamical system with variable $\phi$ that is directly
related to synchrony: (\ref{1d}) relates the parameters $\epsilon$,
$\alpha$, and $\mu$ using $\Omega$ and $H$, as well as a time delay
$\tau$, to the relative phase positions of the two oscillators in
their limit cycle. Now suppose the dynamics $\phi(t)$ from any
starting phase difference $\phi(t_0)$ converge with time to some fixed
phase difference $\phi^*$ with $G(\phi^*)=0$. We refer to this
phenomenon as phase locking, and say that the system is in-phase
synchronized when phase-locked at $\phi^*=0$, and asynchronized when
phase locked at $\phi^*\ne 0$. We also use the term anti-phase synchronization for systems phase locked at $\phi^*=\pi$ specifically.

We solve for $H$ numerically using the numerical software
XPPAUT (Ermentrout 2002). We first find the numerical solution of a
single patch model with given parameters $\epsilon$, $\alpha$, and
$\mu$. We then calculate $\Omega$, and the function $\hat{\gamma}(t)$ that
solves (\ref{adjointz}) (the iPRC, discussed in Appendix~A). XPPAUT can then approximate the
corresponding coupling function $H(x)$ (the calculation uses
(\ref{H})). We obtain this function via its first eleven Fourier
coefficients $a_n$ and $b_n$, $n=0,\ldots,10$, which
approximate
\begin{equation*}
H(x)=\sum_{n=0}^{10} [a_n\cos(nx)+b_n\sin(nx)].
\end{equation*}
Now by a simple calculation (as done in Kobelevskiy (2008)), we define
\begin{align*}
H_{delay}(\phi)&=H(-\phi-\Omega\tau)-H(\phi-\Omega\tau)\\
&=-2\sum_{n=0}^{10}\sin(n\phi)[a_n\sin(n\Omega\tau)+b_n\cos(n\Omega\tau)],
\end{align*}
so that calling $c_n=a_n\sin(n\Omega\tau)+b_n\cos(n\Omega\tau)$, we have an expression for $G(\phi)=\frac{d\phi}{dt}$:
\begin{equation}
G(\phi)=-2\delta\sum_{n=0}^{10} c_n\sin(n\phi).\label{Gg}
\end{equation}
Our approximation was not improved by using more than eleven Fourier
coefficients. We can now integrate $G(\phi)$ in time to find the
solution of the system for an arbitrary starting phase difference to
investigate the effect of $\tau$ on phase dynamics, or simply plot
$G(\phi)$ to determine the phase-locked states and their stability.

\section{Results}\label{results}

In this section we investigate the effect of the time delay $\tau$ on the phase-locked states of the
phase model (\ref{1d}) for one particular set of $\epsilon-\alpha-\mu$ parameters
using the numerical techniques of Section~\ref{analysis}. A validation of these results against
the dynamics of the full model (\ref{full}) is presented in Appendix~B, while the effect of
time delay for a range of $\epsilon-\alpha-\mu$ parameters is considered in Appendix~C.

\subsection{Phase-locked states}

We begin by investigating the effects of a time delay for the
parameter set $\epsilon=0.1$, $\alpha=0.35$, and $\mu=0.3$ in (\ref{nondim}), which for
an uncoupled patch (i.e., $\delta=0$) leads to stable oscillatory dynamics with frequency
$\Omega=0.856$. We use $\delta=0.001$ for weak coupling. We retrieve
the coefficients $a_n$ and $b_n$ to construct
$G(\phi)=\frac{d\phi}{dt}$ from (\ref{Gg}), which has roots corresponding to steady
states. Notice that $G(\phi)$ is $T$-periodic in $\tau$. Thus we
examine a range of $\tau$ values from $0$ to $T=7.34=2\pi/\Omega$. This range is
justified under the assumption that $\tau$ is of the same order of
magnitude as $T$ or less. If $G'(\phi^*)<0$ at a steady-state
$\phi^*$, then it is stable, while if $G'(\phi^*)>0$ then the steady
state is unstable.

\begin{figure}
    \centering
    \includegraphics[width=129mm]{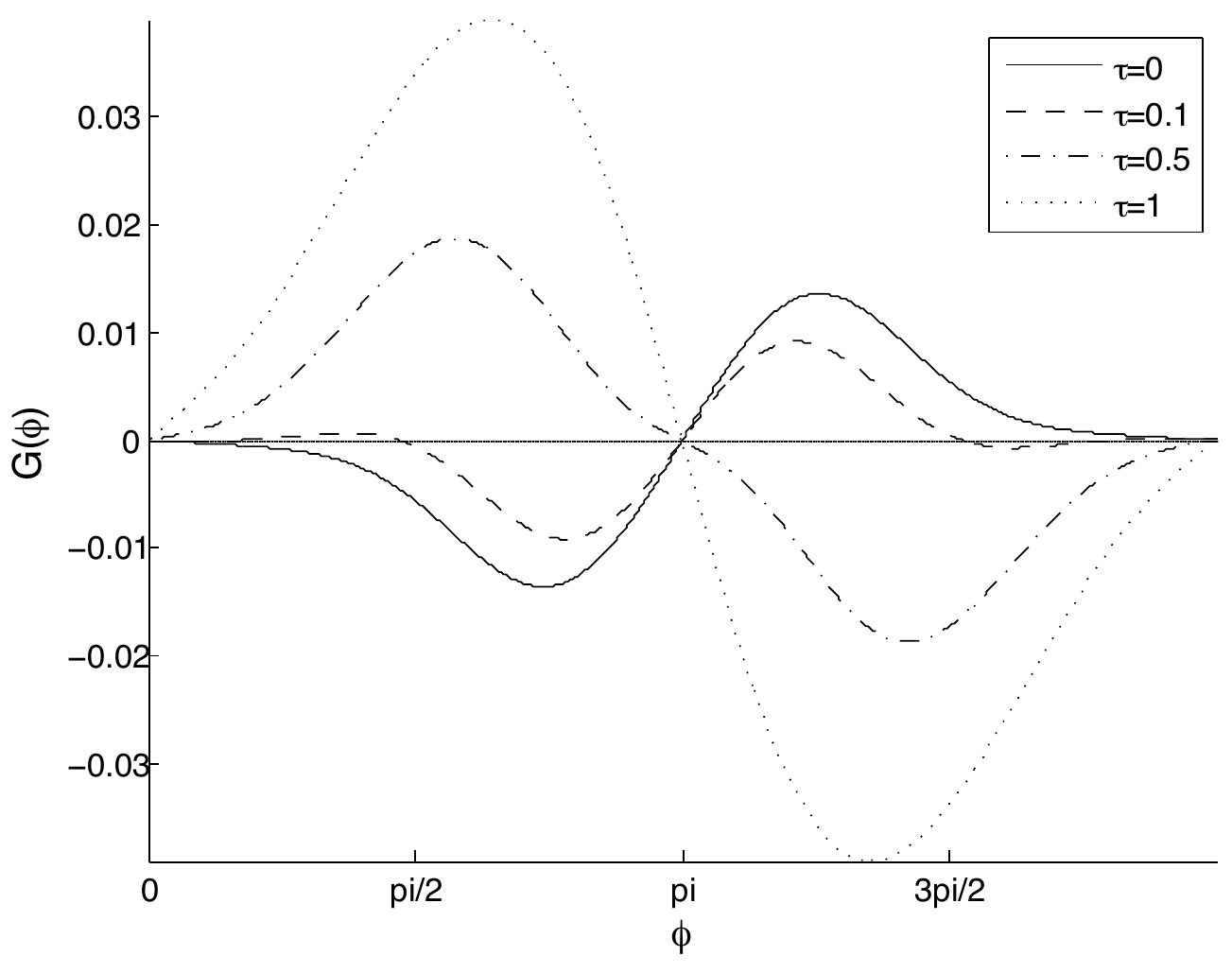}
\caption{$\epsilon=0.1$, $\alpha=0.35$, $\mu=0.3$, and $\delta=0.001$. $G(\phi)$ for a few different values of $\tau$. }
    \label{G}
\end{figure}

\begin{figure}
\centering
\subfigure[]{\includegraphics[width=75mm]{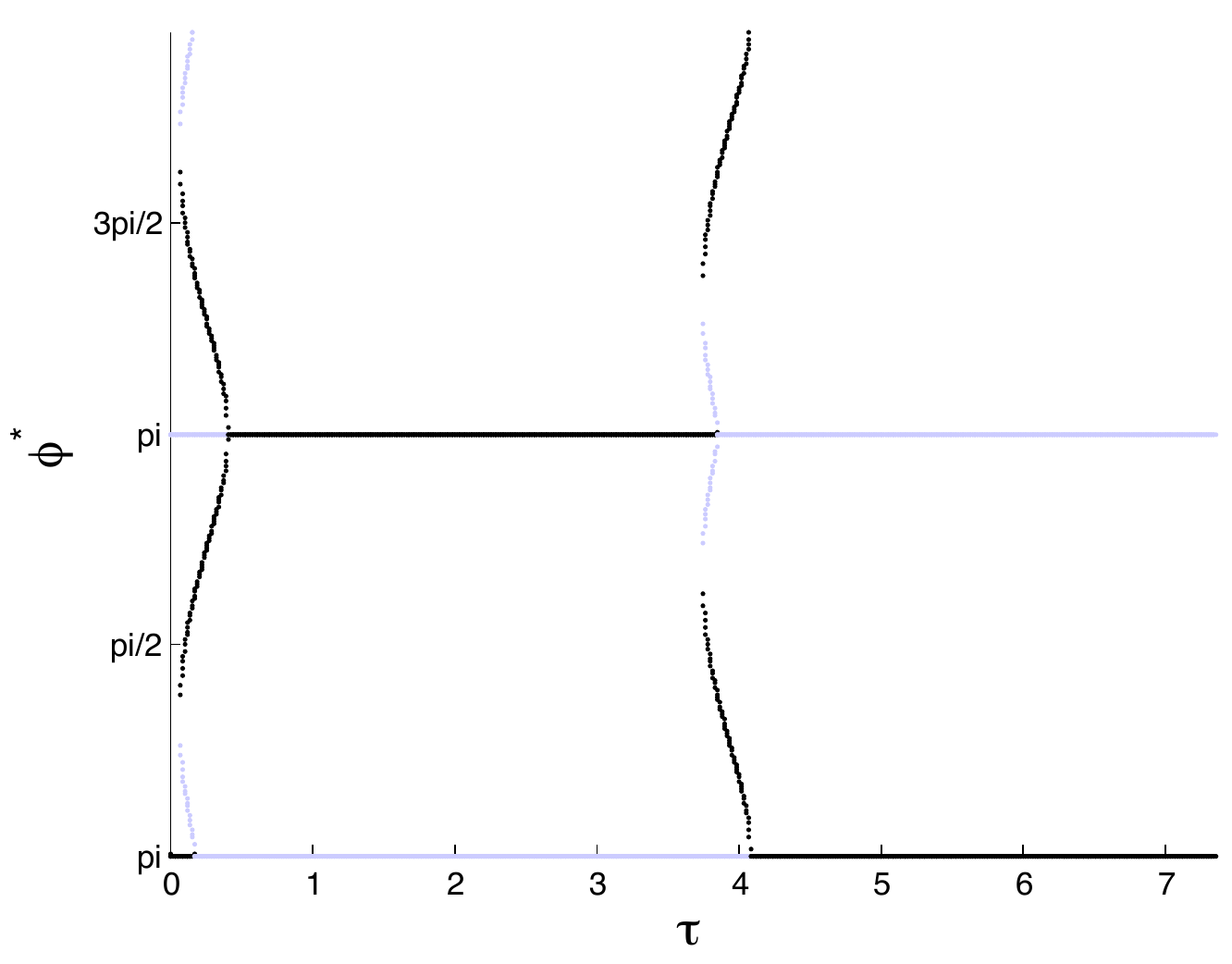}}
\subfigure[]{\includegraphics[width=75mm]{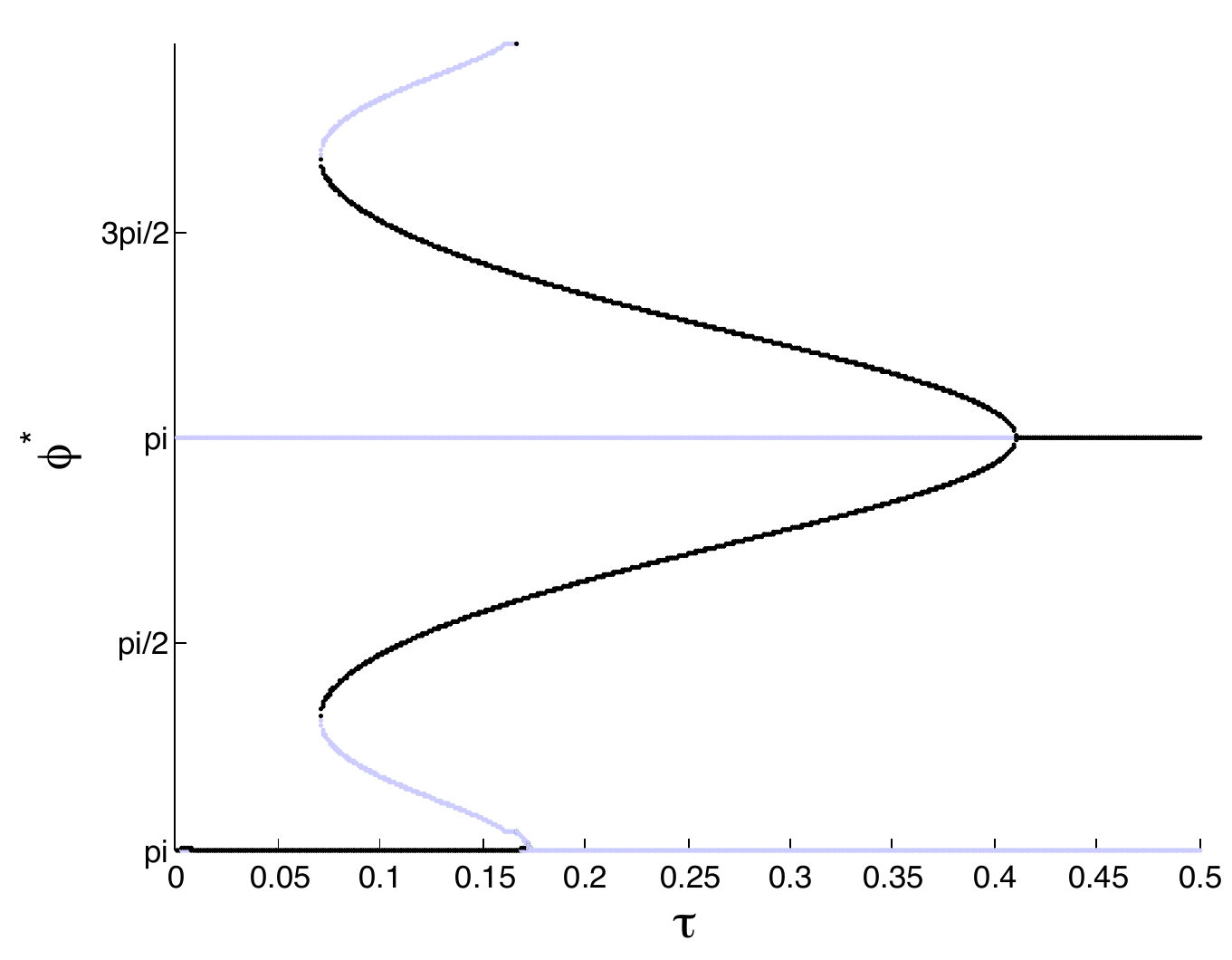}}
%\subfigure[]{\includegraphics[width=64.5mm]{Fig2b}}

\caption{$\epsilon=0.1$, $\alpha=0.35$, $\mu=0.3$, and $\delta=0.001$. $\tau$-Bifurcation diagram: (a) showing a full $\tau$ range  (i.e., over a full period of oscillation),
(b) showing values for $\tau$ close to 0. Stable steady-states are black; unstable steady-states
are blue (appearing as grey in monochrome).
}
\label{bifdiags}
\end{figure}

The number and stability of phase-locked solutions to the two-patch
predator-prey system is strongly affected by delayed dispersal (Figure
\ref{bifdiags}). We start with $\tau=0$, equivalent to instantaneous travel between
patches, and see that the system has one stable steady state solution,
$\phi^*=0$, and one unstable steady-state solution, at $\phi^*=\pi$ (Figures~\ref{G} and~\ref{bifdiags}). As $\tau$ is increased, $0$
and $\pi$ are always steady-state solutions because $G(\phi)$ is
$2\pi$ periodic and odd. We see that other steady-state solutions may
exist, appearing or disappearing in a total of eight saddle-node or
pitchfork bifurcations as $\tau$ is varied from $0$ to $T=7.34$
(Figure~\ref{bifdiags}).

Our bifurcation analysis reveals the complex response of phase dynamics
to a time delay in dispersal.
Super-critical pitchfork bifurcations occur at
the steady state $\phi^*=\pi$ when $\tau=0.41$,
and at the steady state $\phi^*=0$ when $\tau=4.08$.
In the bifurcation at $\tau=4.08$, as can be seen in Figure~\ref{bifdiags}(a),
the steady state at $\phi^*=0$ changes stability,
and a branch of stable steady states bifurcates from $\phi^*=0$
with the new stable branch existing for $\tau<4.08$ where the $\phi^*=0$
solution is unstable. There is thus a stable steady state at $\phi^*=0$
on one side of the bifurcation, and a stable steady state close to $0$
on the other side of the bifurcation. Thus at this bifurcation
the $\phi^*=0$ stable steady-state is carried
continuously away from $0$ (or the reverse). The behaviour at the other supercritical
pitchfork bifurcation at $\phi^*=\pi$ when $\tau=0.41$ is similar, except we
see that there are actually two branches that bifurcate from $\phi^*=\pi$ in symmetric
fashion. This is characteristic of all supercritical pitchfork bifurcations (and also
occurs in the bifurcation at $\phi^*=0$, $\tau=4.08$ on noting that $\phi=0$ and $\phi=2\pi$
represent the same phase).

Sub-critical pitchfork
bifurcations occur at $\phi^*=0$ when $\tau=0.17$
and at $\phi^*=\pi$ when $\tau=3.84$ (see Figure~\ref{bifdiags}).
These are the most dynamically disruptive type of pitchfork bifurcation, since
a stable steady state exists only on
one side of these bifurcation points. A system in nature
passing through such a bifurcation would `jump' abruptly from one steady-state to
another (outside the region of the bifurcation point).

Saddle-node bifurcations correspond to two
steady states (one stable and one unstable) that collide and annihilate so that a stable steady
state either appears or disappears. There are four such bifurcations in our system: two at $\tau=0.07$ ($\phi^*=1.00$ and $5.28$) close to the subcritical pitchfork bifurcation at $\tau=0.17$, and two at $\tau=3.74$ ($\phi^*=2.14$ and $4.15$), close to the subcritical pitchfork bifurcation at $\tau=3.84$.

The fact that there are eight bifurcations across the range of $\tau$
where the stable steady states change indicates that the qualitative
dynamics of the weakly-connected system, and therefore its ability to
synchronize, is remarkably sensitive to the value of the time delay
$\tau$.
Increasing $\tau$ from $0$ to $T$ can produce a full range (i.e.,
from $0$ to $2\pi$) of stable phase differences $\phi^*$ (Table~\ref{steadystates}).
This enriches our understanding and ability to
predict the dynamics of the system beyond the $\tau=0$ case with only one stable steady state ($0$). When $\tau>0$ it is still possible to have
$0$ as a stable steady state, but other
asynchronous stable steady states are available as well over some values of $\tau$, including anti-phase synchronization at phase difference $\pi$.
However, there are also ranges of $\tau$ values where the steady state $0$ becomes unstable, and the system is unable to synchronize in-phase (i.e., for the entire interval between $\tau=0.17$ and $\tau=4.08$). Overall, our
bifurcation diagram (Figure~\ref{bifdiags}) can be used to show that the fraction
of $\tau$ values (of the full range length $T$) where in-phase
synchronization is possible (i.e., $0$ is a stable steady-state) is
$(T+0.17-4.08)/T=0.47$. This means
that over half of all possible time delays (on the same order
of magnitude as $T$ or less) prevent in-phase synchronization given
the parameter values we used in numerical simulations.

In fact, the system loses its ability to synchronize in-phase (i.e.,
the steady state 0 becomes unstable) even with a very small time
delay. If $\tau$ were increased from 0 to just 0.17 (a time delay of
$0.17/\alpha$ in unscaled time), a population with in-phase
synchronized dynamics would bifurcate abruptly (through a sub-critical pitchfork bifurcation) to a significantly asynchronous non-zero stable steady
state, decreasing the risk of simultaneous extinction in both
patches. The phase difference would remain held away from in-phase
synchronization as $\tau$ is increased, all the way until
$\tau=4.08$. This sensitivity of the dynamics to small changes of
$\tau$ also means that if two patches are synchronized in-phase, their global
(simultaneous) risk of extinction can be reduced by a change in the delay $\tau$
of less than $(T+0.17-4.08)/2=1.72$
to bring the dynamics to a sub-critical
pitchfork bifurcation where the system shifts to another non-zero
steady state at $T+0.17$ (recalling that the bifurcation diagrams are T-periodic), or to a super-critical pitchfork bifurcation where the
steady-state phase difference smoothly increases from zero at $\tau=4.08$.

\begin{table}
\caption{$\epsilon=0.1$, $\alpha=0.35$, $\mu=0.3$, and $\delta=0.001$. Stable steady states for varying $\tau$ values (i.e., on the interval from $0$ to a full period of oscillation).}
\label{steadystates}
\begin{tabular}{lll}
\hline\noalign{\smallskip}
$\tau$ Range & Stable steady-states  \\
\noalign{\smallskip}\hline\noalign{\smallskip}
(0, 0.07)& 0 \\
(0.07, 0.17) & 0, $0<\phi^*<\pi$, -$\phi^*$\\
(0.17, 0.41)& $0<\phi^*<\pi$, -$\phi^*$ \\
(0.41, 3.74)& $\pi$\\
(3.74, 3.84)& $0<\phi^*<\pi$, $\pi$, $-\phi^*$\\
(3.84, 4.08)& $0<\phi^*<\pi$, -$\phi^*$\\
(4.08, 7.34)& 0 \\
\noalign{\smallskip}\hline
\end{tabular}
\end{table}

\subsection{Rate of convergence to synchrony}

Previous studies by Goldwyn and Hastings (2008) have shown that in general, relaxation oscillators converge to
steady-state dynamics faster than other types of oscillators (due to their iPRCs having greater maximum magnitudes). We find that within this category of relaxation oscillators, a time delay plays a significant role in further determining how strong a synchronizing force weak dispersal can be. We find that convergence strength to synchrony predicted by the model varies widely over the range of $\tau$ values we
consider, and in a way that indicates that the system is highly
sensitive to small changes of $\tau$.

The local convergence rate to synchrony of a stable steady-state $\phi^*$ is
calculated as $|G'(\phi^*)|$, the absolute value of the first derivative with respect
to $\phi$ of the function $G(\phi)$, evaluated at $\phi^*$; or in
other words, the absolute value of the slope of the graph of $G(\phi)$
at the zero $\phi^*$ (the slope will be negative if $\phi^*$ is a
stable steady-state). The greater the local convergence rate to synchrony,
the shorter the time for transient dynamics around the steady-state;
and the more likely it is to be observed in natural systems exposed to
recurring perturbations away from their phase-locked state.

\begin{figure}
\centering
%\subfigure[]{\scalebox{1}{\includegraphics{Fig3}}}
\subfigure[]{\includegraphics[width=129mm]{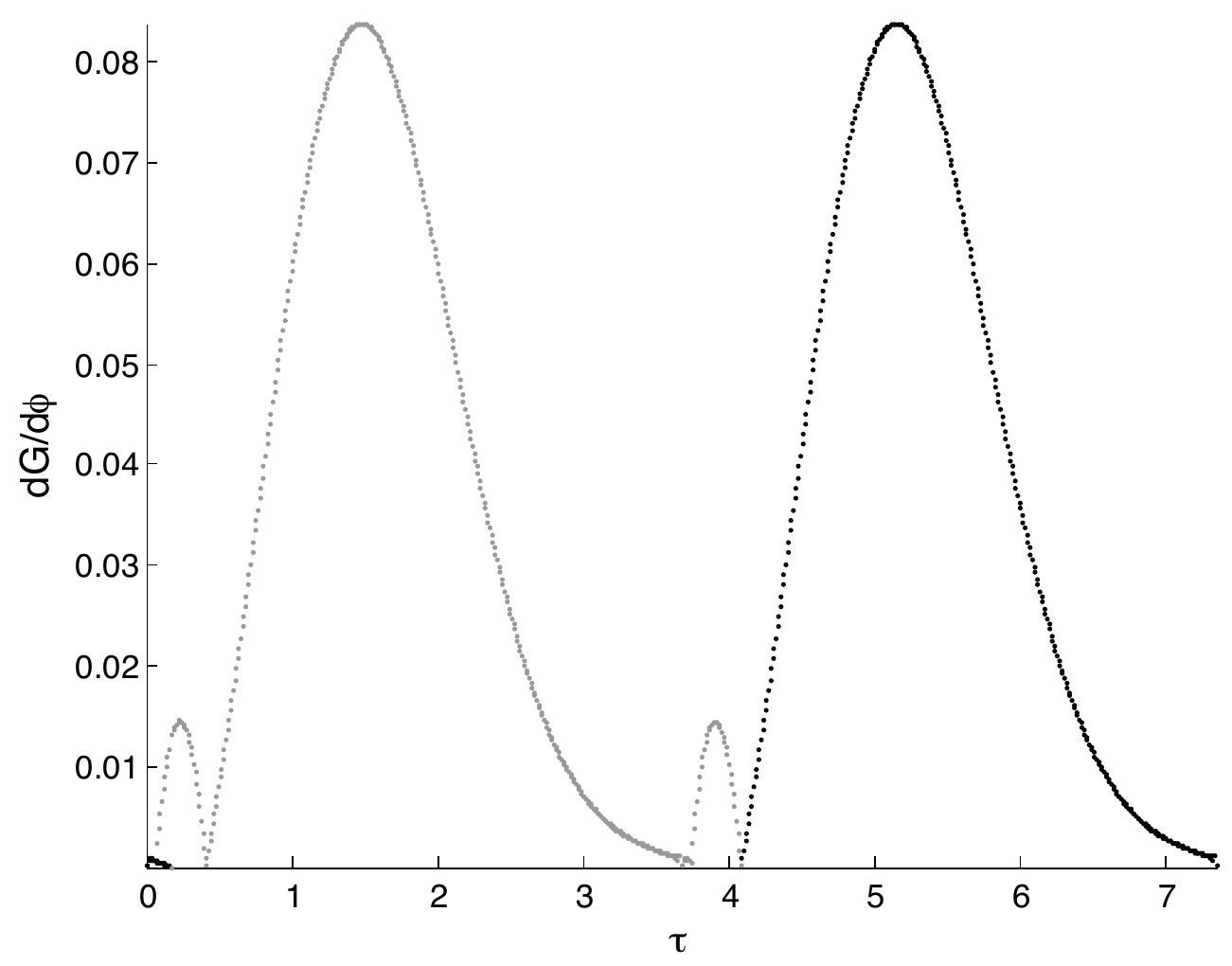}}
\caption{$\epsilon=0.1$, $\alpha=0.35$, $\mu=0.3$, and $\delta=0.001$. Rate of convergence to stable steady-state $0$ (black) or the next stable steady state $\phi^*>0$ (gray) if $0$ is unstable. The values are shown for the full $\tau$ range (i.e., over a full period of oscillation).}
\label{convergeboth}
\end{figure}

Our analysis shows that weak, non-delayed dispersal is not in
itself a strong mechanism of in-phase synchrony in terms of convergence strength, compared to delayed dispersal. This is revealed by the slow convergence to in-phase synchrony at $\tau=0$
compared to those systems with delayed dispersal ($\tau>0$) that have
an in-phase stable equilibrium ({\it e.g.} $\tau=5.5$;
Figure~\ref{convergeboth}).  More specifically, the rate of convergence to a stable
$\phi^*=0$, or to the steady state with the smallest phase difference,
is significantly faster with than without a time delay (Figure~\ref{convergeboth}). This is because
at $\tau=0$ the system is close to the pitchfork bifurcation observed
at $\tau=0.17$ where $0$ loses its stability, and it is precisely
at that bifurcation points that $G'(\phi)=0$ and that the convergence
rate is minimum. Local convergence rate is fastest at the centres of the
intervals between consecutive bifurcations points: these are the
$\tau$ values where weak dispersal is able to best produce
synchronized dynamics in nature. Because there is a bifurcation
very close to $\tau=0$, a slight increase in $\tau$ from zero not only
forces the system into its asynchronous and eventually anti-phase synchronous phase-locking possibilities, but also increases
the rate of convergence to these phase-locked states, and thus the
predicted ability of weak dispersal to explain synchrony in
nature. Thus, we have shown that for a specific set of parameters
corresponding to oscillatory dynamics of a single predator-prey patch,
even a very small time delay can produce dynamics that are markedly
different from the predictions of the non-delayed model.

\section{Discussion}

The importance of a landscape matrix in controlling movement of
individuals and propagules between discrete populations has led to its integration into recent management and conservation practices, but most
theoretical frameworks of patch dynamics are still based on
instantaneous dispersal. Our time-delay approach to modeling dispersal
in a patch model constitutes a simple way to incorporate details of
individual movement through the landscape matrix within a patch
dynamics framework. We show that our patch model with delayed
dispersal provides a useful approximation of dispersal through
continuous landscapes. Our results show that dispersal time is a spatial process deserving of attention when considering
synchronization dynamics. We more precisely show how delayed dispersal
can greatly limit the ability of weak dispersal to synchronize
predator-prey dynamics to in-phase dynamics, which put a system at highest risk of global extinction. Instead, delayed
dispersal is predicted to promote asynchronous phase-locked dynamics across
discrete populations. Our results have important implications for
understanding causes of spatial synchrony in natural systems and for
bridging gaps between simple patch models and landscape ecology.

\subsection{Knowing when to be discrete, and when to be continuous}

Patch dynamics with delayed dispersal can provide a very useful tool
to approximate complex movement patterns across natural landscape
matrices. This approach effectively embeds a patch model onto a more
concrete physical space resembling a fragmented landscape, where local
patches are separated by significant space that individuals disperse
across. The modeling technique for dispersal here is applicable to any
such system in nature where patches are not directly next to each
other (in which case $\tau$ would just be zero), even if there's only
a small transmission time between patches: there is no
limit in general to how sensitive a system can be to small transmission delays.

The notion of time delays in dispersal can be extended to systems
involving egg banks, seed banks, dia-pausing eggs, or dormant stage in
general. It is usually thought that a dormant stage is a means of
``dispersal through time'' to increase survival through harsh
conditions (Callaghan and Karlson 2002). However, many examples of dormant stages also
involve spatial aggregation and thus movement between dormant and
non-dormant stages (Rothhaupt 2000). Such situations where dormant propagules disperse
through space are just an extreme example of the complex ways that
dispersal through space and time are coupled and can explain important
ecological phenomena such as persistence and coexistence
through storage effects (Chesson 2000). Our work shows indeed that
spatial segragation comes hand in hand with temporal segregation, which provides a relevant and insightful
approach to modeling population dynamics in spatial systems.

\subsection{Time-delayed dispersal and synchrony}

Identifying causes of synchrony remains a major challenge in many
biological systems. Ecologists have provided both theoretical and empirical evidence for weak dispersal as an
important cause of spatial synchrony between discrete populations (see Liebhold 2004 for a review), but
the question remains: when can weak dispersal be the mechanism behind
in-phase synchronization between natural populations that are typically exposed
to frequent phase perturbations and embedded within complex
landscapes? Goldwyn and Hastings (2008) showed that a requirement for weak
dispersal to cause synchronization in a network of predator-prey
patches in nature is that the predator-prey dynamics behave as a
relaxation oscillator. Relaxation oscillations
result from a separation of temporal scales between predator and prey
dynamics and lead to fast convergence time to in-phase steady state
synchrony. Within this region of parameter space, our study provides a new perspective on the ability of weak dispersal to cause synchronization across natural landscapes where dispersal between habitats is delayed.

Our time-delayed dispersal approach gives a way of using information
we may have about a system connected through weak dispersal (the
distance between patches, speed of the dispersers, etc.) to infer the
importance of weak dispersal as an important factor behind
synchronized dynamics. Our results show that the delay itself can have
a number of implications for phase dynamics: systems may be carried
away from in-phase synchronization and held at a non-zero
phase-locking state. Over a range of delay values ($\tau$) the system may
increase from one to three the number of stable phase-locked steady states, further
reducing the parameter space resulting in in-phase
synchronization. The two patch Rosensweig-MacArthur model studied here has been associated with bifurcation to anti-phase synchrony in relation to delayed coupling (Prasad et al 2008).  Understanding how anti-phase synchrony arises as a mode of dynamics in natural ecological systems is of increasing interest, given its observation in natural systems as a striking phenomenon with significant ecological implications (see Grenfell et al. 2001, He and Stone 2004, Beninca et al. 2009, Koelle and Vandermeer 2004, for examples). Our analysis predicts the general response of phase synchrony (including anti-phase synchrony) to changes in dispersal delay between patches. Our results will contribute to the applicability of phase-dynamical theories (Goldwyn and Hastings 2008) in natural and experimental ecosystems.

Our results also relate the value of $\tau$ to the
rate of convergence to in-phase synchrony and show how a time delay can greatly impede the ability of weak dispersal to explain the
maintenance of in-phase synchrony in nature, and instead enhance asynchronous and anti-phase synchronous dynamics. The rate of convergence
is important because in nature where systems frequently experience
perturbations due to stochastic fluctuations in the environment, their
steady state dynamics are unlikely to be observed unless they
recover quickly relative to the frequency of perturbations. In
agreement with more general bifurcation theories (Strogatz 2000), our
results show that the local convergence rate is minimum near $\tau$
bifurcation thresholds, and is maximum between these bifurcation
points. This is especially relevant because for a broad range of
$\epsilon$--$\alpha$--$\mu$ values (see Appendix~C), in-phase synchrony bifurcates to phase locking at delay values very close to $\tau=0$. Instantaneous dispersal dynamics is thus highly sensitive to the introduction of arbitrarily small
dispersal time.

Our results suggest that the combined knowledge of all stable
phase-locked states and of their convergence strength is
important to predict when weak dispersal can be a very important cause
of synchronization for systems characterized by relaxation
oscillations. An example of such a system are those involving insect outbreaks, which commonly produce large
amplitude oscillations (Peltonen et al. 2002). In communities characterized by small amplitude and sinusoidal oscillations, such as in
the hare-lynx system, weak dispersal is an unlikely cause of
synchronization, and extra information about the transmission delay
may not be useful. Such systems are more likely to be synchronized
through the Moran effect
(Goldwyn and Hastings 2008), or by strong dispersal as dispersal strength decreases convergence
time. (The local convergence rate $|G'(\phi^*)|$ in our model inherits the factor $\delta$ from $G(\phi)$ -- see (\ref{1d})
or (\ref{Gg}) -- and so the local convergence rate is proportional to $\delta$ and the convergence time
is proportional to $1/\delta$.)

\subsection{Patch dynamics within landscape matrices}

Fragmented landscapes are typically formed of habitats, connecting
corridors, and the overall landscape matrix (Turner et al. 2001). Dispersal time
captures a number of landscape properties: spatial arrangement of
habitats, the presence and effectiveness of corridors, and the
resistance of the landscape matrix to movement (Koh et al 2010). The idea of
movement time (time-delay) between habitat fragments as assumed in our
study can inform on the role of habitat corridors in connecting
wildlife refuges. The use of these corridors in fragmented habitats to
assist the dispersal of endangered species and decrease the risk of
regional extinction is still controversial (Brady et al. 2011), and
could be resolved through the understanding of dispersal delays across
landscapes.

While the presence of a landscape matrix and of corridors involve a
finite dispersal time between habitats, it can also affect demographic
processes and behavior of dispersing individuals, as well as regulate the rate of
dispersal and modulate the coupling strength between habitats
(Koh et al 2010). The phase model we adopted allows studying movement time within a patch dynamical framework, but it also assumes weak dispersal
and no change in individual density or behavior during
dispersal. Future work should investigate the importance of a
transmission delay with stronger dispersal coupling using
delay-differential equations (our model validation in Appendix~B outlines one such analytical method). The
integration of demographic processes such as mortality during
dispersal would also improve the relevance of patch dynamical models
for understanding synchrony across complex landscapes. We hope our
modeling approach can integrate these more realistic assumptions
that are key to conservation, while still contributing to a more general theory of patch dynamics.

\begin{acknowledgements}
E.\ Wall is grateful to the McGill University Biology Department for a Science Undergraduate Research Award (SURA). F.\ Guichard and A.R.\ Humphries thank the Natural Sciences and Engineering Research Council of Canada (NSERC) for funding though the Discovery Grants Program.
\end{acknowledgements}

\section*{Appendix A}
\renewcommand{\thesection}{A} \label{theoappend}

For ease of analysis in studying synchronization with the assumption of weak dispersal, we reduce the full
two-patch model (\ref{nondim}) to a phase model using concepts from
weakly connected network theory (see (Hoppensteadt and Izhikevich 1997)). Here we expand on the phase model reduction used in Section~\ref{theo}, to understand how the general principles of weakly connected network theory can be used on a general system viewed from a coupled oscillator perspective. A
weakly connected network in general is any system of the form
\begin{equation}
\frac{dX_i}{dt}=F_i(X_i)+\delta{}W_i(X_1,\ldots,X_n), \quad i=1,\ldots,n, \label{genfullsystem}
\end{equation}
where each $X_i(t)\in \mathbb{R}^m$ and $\delta$ is a small
parameter. (For now we do not consider a time delay in coupling.) We
are interested in the case where each decoupled subsystem
($\delta=0$)
\begin{equation}
\frac{dX_i}{dt}=F_i(X_i), \quad i=1,\ldots,n
\end{equation}
has an exponentially orbitally stable limit cycle attractor
$\gamma_i\subset\mathbb{R}^m$.

The limit cycle $\gamma_i$ being a periodic orbit of the system, has
an associated period $T_i$ and a natural frequency
$\Omega_i=2\pi/T_i$. As a closed curve through $m$-dimensional space, each $\gamma_i$ can be parameterized with a phase
variable $\theta_i$ that increases by $2\pi$ in one period. With an
arbitrary starting point $p_i\in\gamma_i$, we define a mapping
$P_i:[0,2\pi) \rightarrow \mathbb{R}^m$
that takes $\theta_i\in[0,2\pi)$ to
the unique corresponding point on $\gamma_i$ that is
$P_i(\theta_i)=Y_i(\theta_i/\Omega_i)=Y_i(\theta_iT_i/2\pi)$, where
    $Y_i(t)\in\mathbb{R}^m$ solves $dY_i(t)/dt=F_i(Y_i(t))$ with
    $Y_i(0)=p$ ($Y_i(t)$ is on the limit cycle $\gamma_i$)
    (Hoppenseadt and Izhikevich 1997).

The fact that the $\gamma_i$ are
exponentially orbitally stable limit cycle attractors means that for
small $\delta$, coupling between the $X_i$ only significantly affects
the phase variables $\theta_i$. For systems where oscillators have
identical frequencies $\Omega_1=\cdots=\Omega_n=\Omega$, Malkin's
Theorem says that solutions of the system (\ref{genfullsystem}) can be
continuously mapped to solutions of the following canonical phase
model defined on the $n$-torus
$\mathbb{T}^n=\mathbb{S}^1\times\cdots\times\mathbb{S}^1$
(Hoppensteadt and Izhikevich 1997):
\begin{equation}
\frac{d\theta_i}{dt}=\Omega+\delta{}H_i(\theta_1-\theta_i,\ldots,\theta_n-\theta_i)+\mathcal{O}(\delta^2), \quad i=1,\ldots,n,
\end{equation}
where $H_i$ are phase coupling functions,
\begin{equation}
H_i(\theta_1-\theta_i,\ldots,\theta_n-\theta_i)=\frac{1}{T}\int_0^T\hat{\gamma_i}(t)\cdot W_i(\gamma_1(t+(\theta_1-\theta_i)/\Omega),\ldots,\gamma_n(t+(\theta_n-\theta_i)/\Omega)) dt \label{Hi}
\end{equation}
and $\hat{\gamma_i}(t)$ solves the system
\begin{align*}
\frac{d\hat{\gamma_i}(t)}{dt}=&-DF(\gamma_i(t))^T \hat{\gamma_i}(t) \\
\hat{\gamma_i}(t)&\cdot\gamma_i{'}(t)=1.
\end{align*}

The function $\hat{\gamma_i}(t)$ is referred to as the infinitesimal
phase response curve (iPRC) (Winfree 1980; Kuramoto 1984).
It measures the degree to which an arbitrarily short and infinitesimally
small perturbation advances (positive valued) or slows (negative
valued) the phase (Goldwyn and Hastings 2008). The iPRC can be thought of as a
measure of the sensitivity of the oscillator to perturbations at each
time $t$ in $[0,T_i)$: at times when the oscillator is affected most
  by perturbations (from dispersal), the iPRC has a greater maximum
  magnitude. Goldwyn and Hastings (2008) show that the
  greater the separation in predator-prey timescales ({\it i.e.},
  the more relaxation-like the oscillator), the greater the sensitivity of the
  oscillator to perturbations, and the higher the maximum magnitude of
  the prey component in the iPRC (Table \ref{adjoints}). This explains
  why relaxation oscillators converge faster to synchrony than more
  sinusoidal oscillators. It also underlies the requirement for
  separated timescales between predator and prey for weak coupling to
  result in synchronization in natural systems (Goldwyn and Hastings 2008).

Crucially, phase model reduction is also possible for weakly connected systems
involving an explicit dispersal delay. The phase model for such
systems is derived in (Hoppensteadt and Izhikevich 1997) and
(Ermentrout 1994). For weak coupling and $\mathcal{O}(1)$ delays
(i.e., delays of the same order of magnitude as the period of
oscillation or less), the delays do not explicitly appear in the phase
model, but rather result in an additional phase shift. The main
theorem of weakly connected delayed systems is stated as follows,
adapted from (Hoppensteadt and Izhikevich 1997) and (Izhikevich 2008):

 \vspace{5mm}\noindent{\bf Theorem (Phase model with delayed coupling):}
 {\it
Consider a weakly connected oscillatory network that has an explicit transmission delay, described by the system
\begin{equation}
\frac{dX_i}{dt}=F_i(X_i)+\delta{}W_i(X_1(t-\eta_{i1}),\ldots,X_n(t-\eta_{in});\,\,\, X_i\in\mathbb{R}^m,
\quad i=1,\ldots,n,
\end{equation}
where the $\eta_{ij}$ are finite nonnegative real numbers, all $\mathcal{O}(1)$. Suppose that each uncoupled system has an exponentially orbitally stable $T$-periodic limit cycle solution $\gamma_i\subset\mathbb{R}^m$. Then, the system can be reduced to the phase model
\begin{equation}
\frac{d\theta_i}{dt}=\Omega+\delta{}H_i(\theta_1-\theta_i-\Phi_{1i},\ldots,\theta_n-\theta_i-\Phi_{1i})
+\mathcal{O}(\delta^2), \quad i=1,\ldots,n,
\end{equation}
where $\Phi_{ji}=\Omega\eta_{ji} \mbox{ mod }2\pi$, and the $H_i$ functions are defined by (\ref{Hi}).

}

\section*{Appendix B} \label{appB}
\renewcommand{\thesection}{B}

\begin{figure}
\centering
\subfigure[]{\includegraphics[width=75mm]{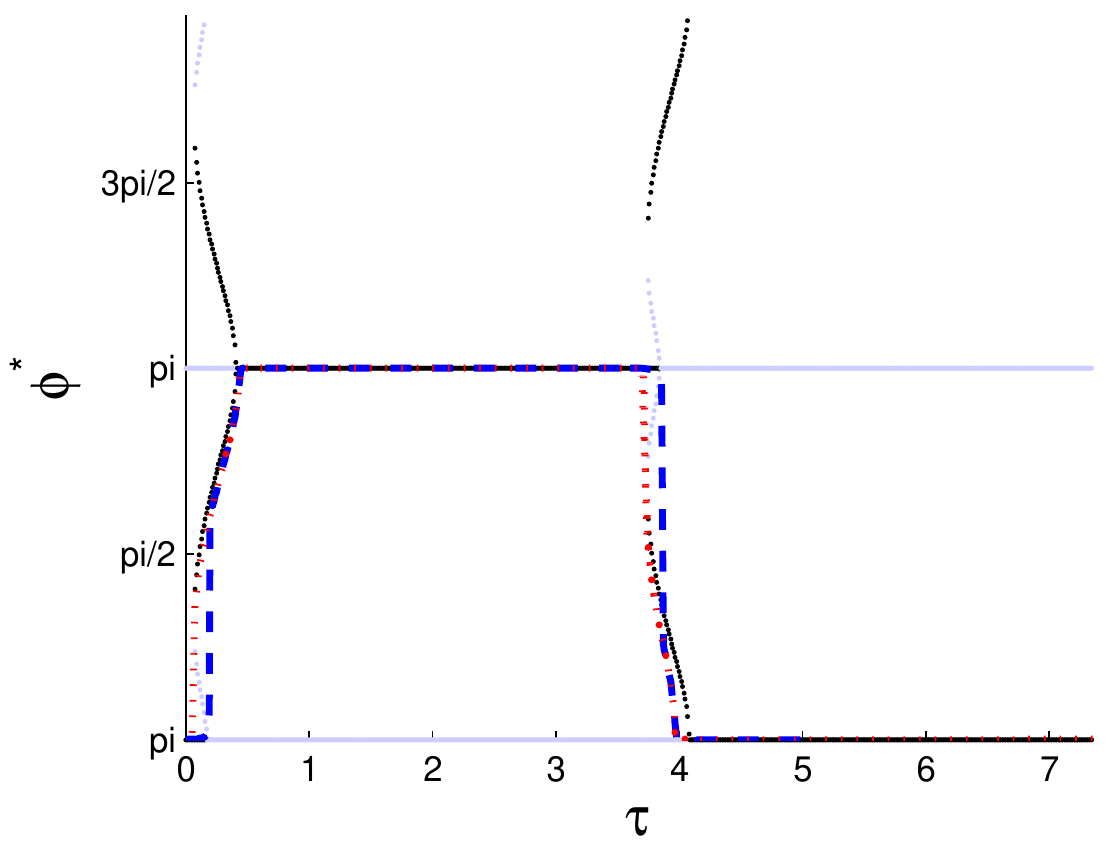}}
\subfigure[]{\includegraphics[width=75mm]{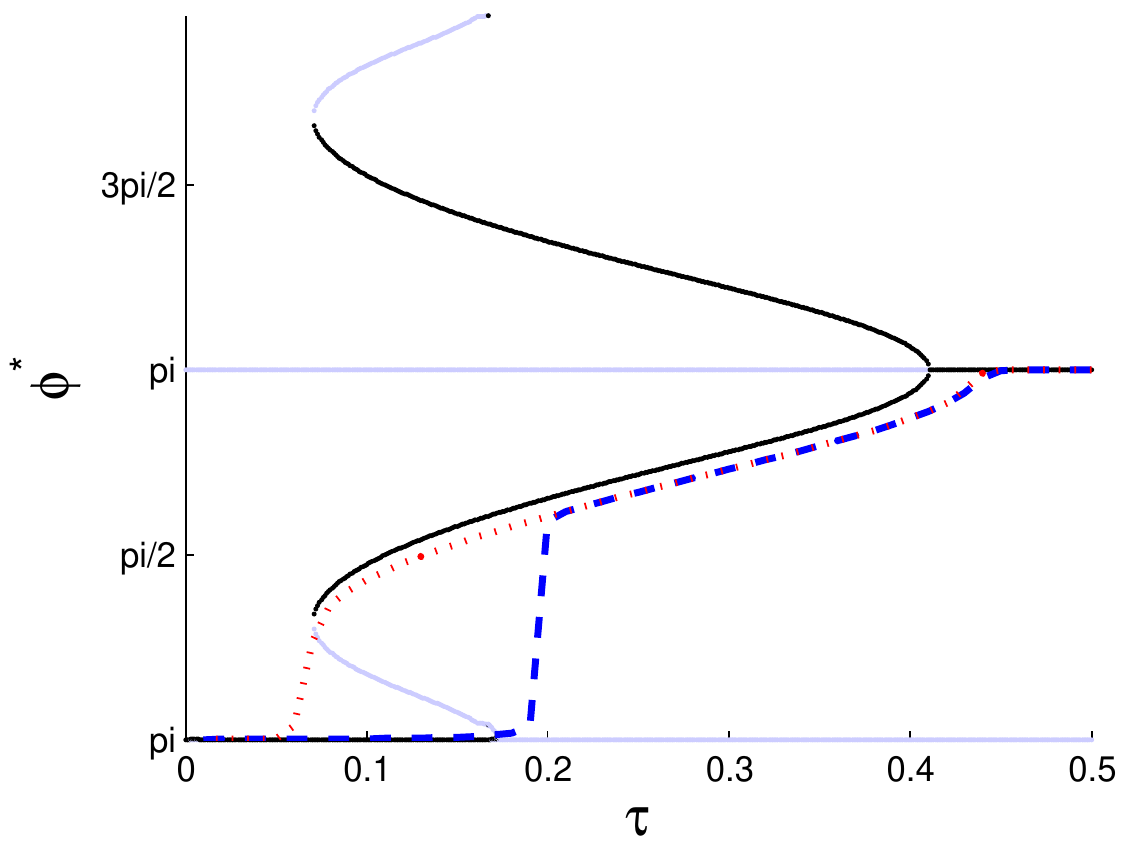}}
%\subfigure[]{\includegraphics[width=75mm]{Fig2b}}
%\subfigure[]{\includegraphics[width=64.5mm]{Fig2b}}

\caption{The $\tau$-Bifurcation diagram for $\epsilon=0.1$, $\alpha=0.35$, $\mu=0.3$,
for the phase reduced model (\ref{1d}) as in Figure~\protect\ref{bifdiags} with superimposed numerically
computed phase differences for the full model (\ref{full}) with $\delta=0.001$ for sequences of
increasing (blue dashed line) and decreasing (red dotted line) $\tau$ values.}
\label{figvalid}
\end{figure}

According to the phase reduction theory of Appendix~A
the bifurcation diagram for the reduced system (\ref{1d}) in Figure~\ref{bifdiags}
should correspond to the phase-locked states of the full system (\ref{full})
in the limit as $\delta\to0$. To validate the theory and our numerical implementation of it,
we also integrated the full system (\ref{full}) using the matlab \cite{matlab} initial value
problem delay-differential equation solver dde23. The results for the same parameter
values as in Section~\ref{results} are shown in Figure~\ref{figvalid}.

We measured the asymptotic phase difference for the full model (\ref{full}) by integrating
the delay-differential equation over a long time interval, then
measuring the asymptotic period of the orbit and phase difference using a Poincar\'e section to
detect successive points where $p_i(t)=1$ with $\dot{p}_i(t)<0$. These long initial value computations were
then repeated over a sequence of $\tau$ values, using the orbit found for one $\tau$ value as the starting
point for the transient integration for the next $\tau$ value. This approach is similar to observing the dynamics
as the delay $\tau$ is slowly adjusted, and different dynamics can be and are observed depending on whether
$\tau$ is increased or decreased.

We found that as $\tau$ is varied
the period of the full system (\ref{full}) could vary by up to 4\% from the period of
single patch model (\ref{single}) with the same parameter values. From Figure~\ref{figvalid}(a) we see that the
asymptotic phase difference is identical for (\ref{full}) and reduced models (\ref{1d}) in the intervals
where the in-phase or anti-phase solutions are stable. From Figure~\ref{figvalid}(b) we see that
when the reduced model indicates a stable phase-locked solution with phase $\phi^*\in(0,\pi)$
then the full model (\ref{full}) has phase locked solutions with slightly perturbed phase. The perturbation
in the asymptotic phase is dependent on the parameter $\delta$ with the results in Figure~\protect\ref{bifdiags}
for the reduced model (\ref{1d}) representing the limiting behaviour as $\tau\to0$. Consequently, for $\delta>0$
the bifurcations at the ends of these intervals are also perturbed slightly. However, it is very hard
to quantify exactly how much these bifurcations are perturbed, since at the bifurcation points the convergence
rate to the asymptotic phase is zero and it would be necessary to integrate through an infinite time transient
to obtain convergence. This is particularly apparent in Figure~\ref{figvalid}(b)
near the bifurcation at $\phi^*=0$ and $\tau=0.17$, where the computed solution for the full model initially
only lifts away slowly from the unstable solution $\phi^*=0$ for $\tau>0.17$.

Figure~\ref{figvalid}(b) also displays the ecologically relevant phenomenon of hysteresis. When
increasing $\tau$ in small increments the solution of the full model stays at the ecologically dangerous phase
locked solution $\phi^*=0$ for all $\tau<0.17$, until this solution loses stability in the subcritical
pitchfork bifurcation at $\tau=0.17$, whereafter for $\tau>0$ the solution rapidly converges to the other branch of
stable phase-locked solutions with $\phi^*>0$. On the other hand, if instead we decrease $\tau$ after the
anti-phase solution loses stability near $\tau=0.41$ the solution remains in a phase locked state with $\phi^*>0$
all the way down to $\tau\approx0.07$ when this solution is destroyed in a saddle node bifurcation, and the phase
rapidly converges to $\phi^*=0$. Thus for $\tau\in(0.07,0.17)$ different dynamics are observed for the full
system depending on whether $\tau$ is being increased or decreased. But this could already have been inferred from
Figure~\ref{bifdiags}, and is related to the existence of the branch of unstable phase locked orbits seen for
$\tau\in(0.07,0.17)$ in the reduced system, and which also exist in the full system
but cannot be found by integration, since they are unstable.

We also validated the numerics with other parameter values and found broadly similar results. As we expected
the less relaxation-like the periodic orbit was (the lower the value of iPRC in Table 2) the longer became
the transient time that we needed to integrate the full system through to see the asymptotic behaviour.
On the other hand, for more relaxation-like oscillators the more sensitive the dynamics became to the
coupling parameter $\delta$.

\section*{Appendix C} \label{appC}
\renewcommand{\thesection}{C}

\begin{figure}
    \centering
%    \scalebox{1}{\includegraphics{bigtau}}
%    \includegraphics[width=129mm]{Fig5}
    \includegraphics[width=150mm]{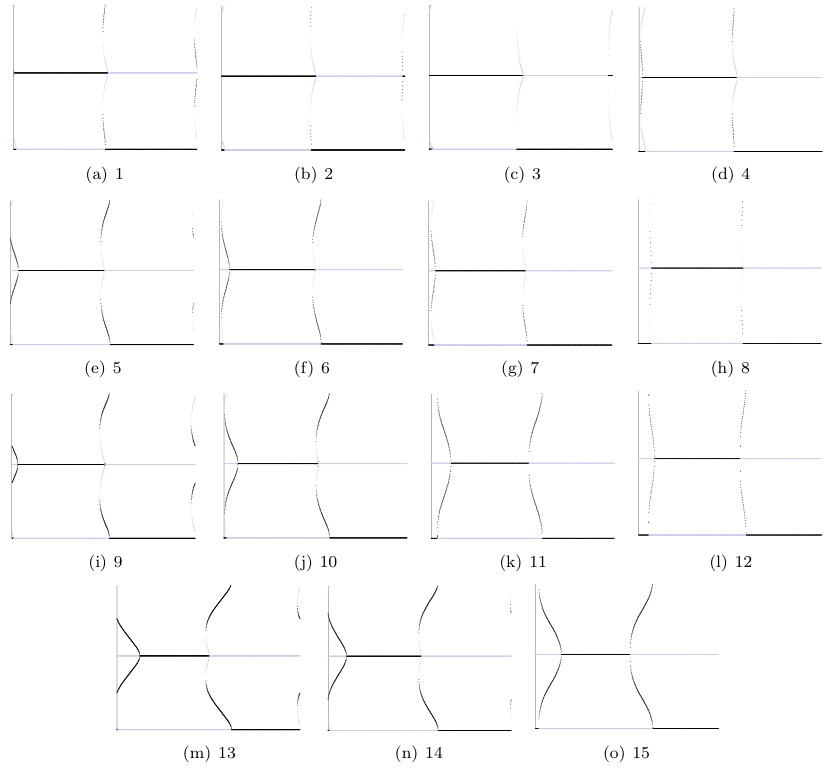}
\caption{$\delta=0.001$; refer to Table \ref{adjoints} for parameter set values corresponding to numbers 1 through 15. $\tau$ Bifurcation diagrams (full range of $\tau$ from 0 to $T$) through a region of $\epsilon-\alpha-\mu$ parameter space. Stable steady-states are black; unstable steady-states are blue.}
\label{more1}
\end{figure}

\begin{figure}
    \centering
%    \scalebox{1}{\includegraphics{smalltau}}
%    \includegraphics[width=129mm]{Fig6}
    \includegraphics[width=150mm]{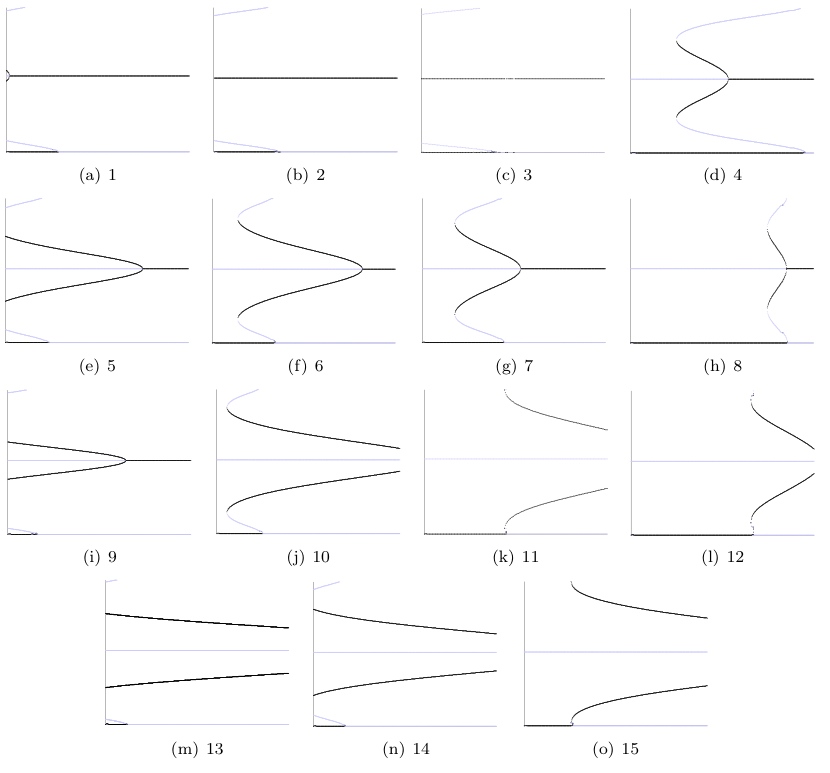}
\caption{$\delta=0.001$; refer to Table \ref{adjoints} for parameter set values corresponding to numbers 1 through 15. $\tau$ Bifurcation diagrams ($\tau$ from 0 to 0.5) through a region of $\epsilon-\alpha-\mu$ parameter space. Stable steady-states are black; unstable steady-states are blue.}
\label{more2}
\end{figure}

We examine specifically how varying the parameters $\epsilon$,
$\alpha$ and $\mu$ might affect the general results we showed for a
specific set of parameters leading to oscillatory dynamics of a single
predator-prey patch. Table \ref{adjoints} shows the parameter sets, as
well as the maximum magnitude of the iPRC at these parameter
values. As expected, decreasing $\alpha$ and $\mu$ (we keep
$\epsilon=0.1$) increases the height of the iPRC. Figures~\ref{more1}~and~\ref{more2} show the
bifurcation diagrams analogous to Figure~\ref{bifdiags}(a) and Figure~\ref{bifdiags}(b) for a number of other
parameter sets. Similarly to the parameter set used for our main
analysis, varying $\tau$ across the interval $[0,T]$ changes the
dynamics and rates of convergence to steady states as the dynamics
bifurcate a number of times. We see that there is a range of parameter space
where $\phi^*=0$ is stable at $\tau=0$, but very close to a
bifurcation, so that with even a slight increase in $\tau$, the system
can move to asynchrony with a much faster convergence
rate (seen best in Figure~\ref{more2}). Investigating this slightly larger region of
$\epsilon$--$\alpha$--$\mu$ phase space indicates that our qualitative
results for the first set of parameters are generally comparable to
other parameter sets, and that decreasing $\alpha$ and $\mu$
(separating the predator-prey timescales and making the oscillator
relaxation-like) mainly has an effect that shows up on the dynamics near
$\tau=0$, and the $\tau$ value near midway through the period of oscillation, where $\phi^*=\pi$ loses its stability.

\begin{table}
\caption{Parameter sets.}
\label{adjoints}
\begin{tabular}{llllll}
\hline\noalign{\smallskip}
Parameter& $\epsilon$ & $\alpha$ & $\mu$ & iPRC (adjoint)\\
set& & & & height\\\noalign{\smallskip}\hline\noalign{\smallskip}
1&0.1&0.35&0.2&1.8$\times 10^4$\\
2&0.1&0.4&0.15&4.5$\times 10^5$\\
3&0.1&0.45&0.1&1.0$\times 10^6$\\
4&0.1&0.45&0.2&180\\
5&0.1&0.3&0.3&1600\\
6&0.1&0.35&0.3&140\\
7&0.1&0.4&0.25&130\\
8&0.1&0.45&0.3&12\\
9&0.1&0.25&0.3&1.0$\times 10^5$\\
10&0.1&0.3&0.35&300\\
11&0.1&0.35&0.4&12\\
12&0.1&0.4&0.35&11\\
13&0.1&0.2&0.45&1.1$\times 10^4$\\
14&0.1&0.25&0.4&900\\
15&0.1&0.3&0.45&20\\

\noalign{\smallskip}\hline
\end{tabular}
\end{table}

\section*{Appendix D} \label{appD}
\renewcommand{\thesection}{D}
\setcounter{equation}{0}

This section is meant to examine in further depth the relationship of our patch dynamics with time
delayed dispersal to other spatial dynamical frameworks. Durrett and
Levin (1994) showed the importance of choosing an approach
to modeling spatially distributed systems-- mean field approaches,
patch models, reaction-diffusion equations, or interacting particle
systems-- that captures the right level of spatial detail for
understanding population dynamics. The time-delay model we consider
(see (\ref{original})) can be viewed as a combination of two existing
treatments of spatial dynamics distinguished in this work. The system
neglecting the time delay (i.e., for $s=0$ in (\ref{original}))
corresponds to the patch model approach: it groups individuals of a
species into discrete patches without any spatially-explicit structure
informing where patches are with respect to one another. In these
models each patch is treated with a mean-field approach where
individuals interact equally with each other and the patch itself is
described by ordinary differential equations, and patches are
connected to all others equally by dispersal or migration at a
constant rate (Durrett and Levin 1994).

Patch models and mean-field approaches both lose the level of spatial
detail required to account for the movement and local interactions of
individuals both in the patch, and in dispersal between patches. The
most common approach to modeling spatial systems working on a finer
level of spatial detail are reaction-diffusion equations, in which
infinitesimal individuals diffuse in space and undergo purely local
interactions (reactions). Allen (1983) shows how the dispersal term in a patch model
(say, $d(h_j(t)-h_i(t))$) can actually be derived from
a simple one-dimensional diffusion equation directly, through the discretization of
continuous space and the assumption of Neumann boundary conditions, which confine the individuals to the union of the finite number of patches.

\begin{figure}
\centering
    \includegraphics[width=129mm]{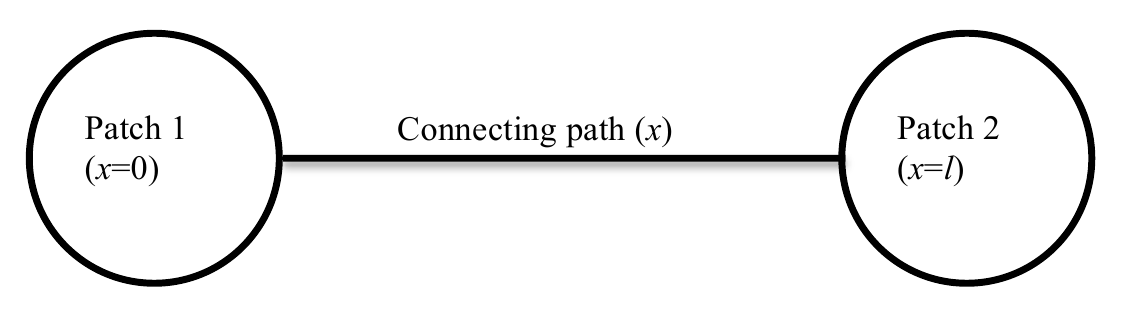}
  \caption{Two discrete (homogeneous) patches connected by a one-dimensional (two-directional) path for dispersal.}
\label{twopatch}
\end{figure}

By treating space as continuous, our time-delay model avoids
discretizing inter-patch space, keeping closer to the
reaction-diffusion equation to describe dispersal in the space between
patches. The addition of continuous space between patches (see Figure~\ref{twopatch})
as opposed to discretizing space allows for handling the concept of
diffusive dispersal closer to the more ecologically realistic
formulation based on finite-speed random walks in the environment
separating the patches, so that dispersers take some meaningful amount
of time to cross from patch to patch. The model retains key features
of actual movement as it makes a random walk between the patches
through the value of the delay $\tau$. Such features are lost in the
discretized version with no delay. Furthermore, our time-delay dispersal is equivalent to the simplest case possible of the process of individuals reaction-diffusing in
space: it assumes a constant mean dispersal duration, and no birth or
mortality during dispersal.

In a system of patches connected by dispersal, the mean transmission
time $\tau$ can be calculated in several ways using specific details
of the individual dispersal process which might be easier to measure
in natural systems than the time delay $\tau$ itself. We now
demonstrate two such derivations. We start with a continuous space
distribution of individuals dispersing in a one-dimensional connecting
path between patches, according to some underlying assumptions about
movement. We then uncover the net squared displacement of individuals
in the path, which give the mean time an individual spends dispersing
between two patches separated by a distance $l$.

If we first assume individuals move passively at a constant speed $v$ with
two-way travel on one-dimensional paths connecting patches, we can
easily see that the mean crossing time $\tau$ in this case is just
$v/l$. Suppose we consider the more general movement pattern integrating
active dispersal modeled by the telegraph equation:
\begin{equation}
\frac{T}{2}\frac{\partial^2 u}{\partial t^2}+\frac{\partial u}{\partial t}=\frac{v^2T}{2}\frac{\partial^2 u}{\partial x^2}, \label{telegraph}
\end{equation}
where $v$ is the finite speed of individual movement, and $T$
is the characteristic time of movement before a direction reversal
(see explanation in Turchin (1998) or derivation in Othmer and Dunbar (1988)). The mean squared displacement over time,
$|x^2(t)|$, of a group of individuals instantaneously released into
the system at patch 1 is calculated as follows in Othmer and Dunbar (1988):
 \begin{equation}
 |x^2(t)|=v^2T\bigl[t-\frac{T}{2}(1-e^{-2t/T})\bigr].
 \end{equation}

While the mean displacement $|x(t)|$ is not quite equal to the square
root of the mean squared displacement, one can scale $\sqrt{|x^2(t)|}$
by some factor $C$ to obtain a better estimate for $|x(t)|$ than
$\sqrt{|x^2(t)|}$ (Byers 2001). Then the mean displacement is then
\begin{equation}
|x(t)|=C\sqrt{v^2T\bigl[t-\frac{T}{2}(1-e^{-2t/T})\bigr]}. \label{disp}
\end{equation}
Now set $|x(t)|$ in (\ref{disp}) to $l$ and solve for the
corresponding value of time $t$. This defines the delay $s$ which
appears in (\ref{original}) as a function of $l$, $T$, and $v$. Note that
the right hand side of (\ref{disp}) is a strictly monotonically
increasing function of $t$, so there is a unique solution.

\end{document}